\documentclass[12pt]{article}

\usepackage{textcomp}
\usepackage{amsmath} 
\usepackage{graphicx} 
\usepackage{setspace} 
\usepackage{authblk}
\usepackage{tabularx} 
\usepackage{booktabs} 
\usepackage{cite}
\usepackage[a4paper,
			bindingoffset=0.2in,
			left=0.5in,
			right=0.5in,
			top=1in,
			bottom=1in,
			footskip=0.25in]{geometry}

\title{Leveraging the Bi$_2$O$_3$--Fe$_2$O$_3$ Phase Diagram to Tailor BiFeO$_3$ Structure and Dielectric Response}

\author[1,2]{Subir Majumder}
\author[1]{Paul Ben Ishai}
\author[2,*]{Gilad Orr} 
\affil[1]{THz and Dielectric Science Lab., Dept. of Physics, Ariel University, Ariel, Israel}
\affil[2]{Crystal Physics Lab., Dept. of Physics, Ariel University, Ariel, Israel}
\affil[*]{Corresponding Author: gilad.orr@ariel.ac.il} 

\begin{document}

\maketitle
	
\begin{abstract}
Advancing the functional performance of bismuth ferrite (BiFeO$_3$) requires precise control over phase stability and microstructure, challenges often complicated by secondary phase formation within the Bi$_2$O$_3$–Fe$_2$O$_3$ system. In this work, we employ a phase-diagram-guided synthesis strategy to clarify the processing–structure–property relationships governing BiFeO$_3$ ceramics. Based on previous reports and our experimental observations, a refined Bi$_2$O$_3$–Fe$_2$O$_3$ phase diagram was constructed identifing the onset of liquid-phase formation near 835$^\circ$C in Bi-rich compositions. Solid-state synthesis using non-stoichiometric precursor ratios (55$\colon$45 and 70$\colon$30 Bi$_2$O$_3$ $\colon$ Fe$_2$O$_3$) reveals that the 55$\colon$45 composition sintered at 775$^\circ$C favours phase-pure rhombohedral $R3c$ BiFeO$_3$ with suppressed sillenite- and mullite-type impurities, lattice contraction, and improved grain uniformity. These structural refinements result in near-Debye dielectric relaxation and a four-orders-of-magnitude enhancement in electrical conductivity relative to lower-temperature or Bi-rich conditions. This work demonstrates the effectiveness of phase-diagram control as a scalable route to tuning dielectric response in BFO-based multiferroics and provides a foundation for processing optimization in complex oxide ceramics.
\end{abstract}

\vspace{10pt}

\section{Introduction}
Multiferroics represent a class of materials distinguished by their ability to exhibit concurrent ferroelectric, ferromagnetic, and often ferroelastic behaviours within a single or composite structure \cite{Landau84,Spaldin2017}. This unique set of properties has garnered increasing attention, propelled by the diverse array of applications they offer. By combining ferroelectric and ferromagnetic characteristics, multiferroics give rise to the intriguing magneto-electric effect (ME) \cite{Varshney2011,Cheng2008}, wherein the interaction between magnetic and electric fields enables manipulation of both magnetic and electrical properties. This phenomenon holds significant promise for practical implementation, including transducers, magnetic field sensors \cite{Talker2022}, memory devices, and photovoltaic systems \cite{Fiebig2002,Hur2004,Wang2003, Hill2002,Borissenko2013Lattice, Catalan2009Physics, Hassan2022Bismuth,Allibe2012Room}.\\

The poster-boy of these materials is $BiFeO_3$ (BFO) which is capable of simultaneously exhibiting ferroelectric and antiferromagnetic properties at room temperature within the same phase \cite{Graf2015Dielectric}. Structurally, BFO adopts a rhombohedral ferroelectric configuration with a perovskite structure ($ABO_3$) at room temperature and is characterized by a notably high ferroelectric Curie temperature ($T_C \sim 830^\circ$C). It has a G-type antiferromagnetic ordering below the Néel temperature ($T_N \sim 370^\circ$C) \cite{Ilic2016Improving}. However, achieving pure single-phase BFO synthesis poses a significant challenge. Synthesis methods often result in a polycrystalline state accompanied by secondary phases such as Mullite-like ($Bi_2Fe_4O_9$) and Sillenite-like ($Bi_{25}FeO_{39}$) structures, which adopt orthorhombic (Pbam) and cubic (I23) symmetries, respectively. Studies by Liu et al. \cite{Liu2015Synthesis} have demonstrated the ubiquitous formation of secondary phases in a range of temperatures ranging from 650$^\circ$C to 850$^\circ$C. Furthermore, Golic et al. \cite{Golic2016Structural}, using various methodologies, have corroborated the appearance of secondary phases simultaneously with BFO phases within the temperature range of 620--770$^\circ$C. \\

\begin{figure}[h!]
	\centering
	\includegraphics[width=0.9\textwidth]{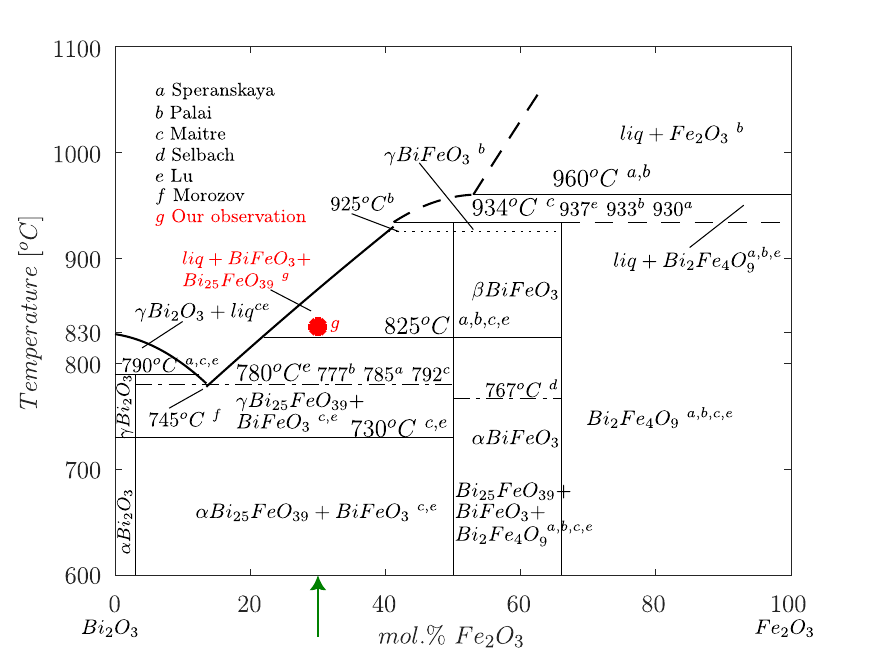}
	\caption{Phase diagram of $Bi_2O_3$--$Fe_2O_3$ system at ambient condition. Contributions from (a) Speranskaya  et al. \cite{Speranskaya1965Phase}, (b) Palai et al.\cite{Palai2008Phase}, (c) Maître et al.\cite{Maitre2004Experimental}, (d) Selbach et al.\cite{Selbach2009On}, (e) Lu et al.\cite{Lu2011Phase}, (f) Morozov et al.\cite{Morozov2003Specific} are indicated along with our observations (g) in red.}
	\label{fig:phase_diagram} 
\end{figure}

The phase diagram of the $Bi_2O_3$--$Fe_2O_3$ system at ambient pressure (Figure \ref{fig:phase_diagram}) was initially mapped in the mid-1960s \cite{Speranskaya1965Phase}, but subsequent revisions have revealed several inconsistencies \cite{Haumont2008Phase, Lu2011Phase, Palai2008Phase, Morozov2003Specific, Maitre2004Experimental}. Significant disagreement persists with respect to the precise temperatures of critical transformations within this system. For example, the onset of the liquid phase has been reported at temperatures ranging from 745$^\circ$C to 792$^\circ$C \cite{Achenbach1967Preparation}. Similarly, the temperature of decomposition of $Bi_2Fe_4O_9$ has been observed to vary between 934$^\circ$C and 961$^\circ$C \cite{Palai2008Phase}, while the peritectic decomposition of $BiFeO_3$ is cited in a wide range, from 852$^\circ$C to 934$^\circ$C \cite{Maitre2004Experimental}.\\

Further complicating matters, Palai et al.\cite{Palai2008Phase} proposed a decomposition pathway for $BiFeO_3$ originating from the $\gamma$ phase, a hypothesis that has been questioned by Lu et al.\cite{Lu2011Phase}, due to the perceived thermodynamic instability of the $\gamma$ phase.\\

To understand these discrepancies, we constructed the phase diagram in Figure \ref{fig:phase_diagram} by compiling data from previous work \cite{Speranskaya1965Phase,Haumont2008Phase,Lu2011Phase, Palai2008Phase,Morozov2003Specific, Maitre2004Experimental, Achenbach1967Preparation} and refining it with our own observations, marked (g) in red. These identify the emergence of a liquid phase at $835^\circ$C for the 70$\colon$30 precursor mixture. This melting behaviour of the bismuth-rich material, which is different from the peritectic decomposition of the stoichiometric $BiFeO_3$ at $\sim$920$^\circ$C (observed further to the right of the diagram), provides a clearer delineation of phase boundaries and improves understanding of this complex system.\\

Various synthesis methods have been employed in attempts to produce BFO, including solid-state reactions \cite{Degues2021Synthesis,Valant2007Peculiarities,Orr2020sintering}, sol-gel processes \cite{GarciaZaleta2014Solid,Sharma2016Magnetic}, and autocombustion methods \cite{Tripathy2013Structural}. However, the precise phase composition in the $Bi_2O_3$--$Fe_2O_3$ system is highly sensitive to small deviations from the ideal stoichiometric ratio \cite{Morozov2003Specific}. For example, a slight deviation to the left of the phase diagram gives sillenite phases, whereas mullite-like phases are likely, whereas the precursor deviates to the right. Although phase-pure $BiFeO_3$ can theoretically be achieved through careful stoichiometric control, this remains challenging due to the sensitivity of phase composition to precursor ratios, synthesis temperature, annealing history and the presence of impurities \cite{Tripathy2013Structural}. Despite extensive efforts to refine the synthesis of pure-phase BFO and explore its thermodynamic behaviour, the phase diagram remains unclear. Although there is general consensus that the $BiFeO_3$ phase is thermodynamically stable over a wide temperature range, conflicting reports persist \cite{Levin1964Polymorphism}. Some studies \cite{Morozov2003Specific, Carvalho2008Synthesis, Selbach2009On} indicate that, under certain conditions $BiFeO_3$ decomposes into secondary phases, challenging both the established phase diagrams and the overall stability of the $BiFeO_3$ phase itself.\\

Our research focuses on the phase diagram (Figure \ref{fig:phase_diagram}) to explore the formation and stability of the BFO phases under varying compositional conditions. The reaction between $Bi_2O_3$ and $Fe_2O_3$  ($Bi_2O_3$ + $Fe_2O_3$ $\longrightarrow$ 2$BiFeO_3$) ideally produces phase-pure BFO in a 50$\colon$50 (mol $\%$) stoichiometric ratio. To investigate the impact of the bismuth content, we adjusted the precursor composition to 55$\colon$45 (mol$\%$) with 10$\%$ excess $Bi_2O_3$ and 70$\colon$30 (mol$\%$) with 40$\%$ excess $Bi_2O_3$ to examine their effects on BFO formation, phase purity, and crystal structure. These variations allow us to systematically study how bismuth enrichment influences structural stability and functional properties such as dielectric behaviour. The insights gained from this study should contribute to the development of more efficient BFO-based materials and to the optimization of their performance for applications in electronics, sensors, and energy devices. Furthermore, this research lays the foundation for the growth of large single crystals of BFO, advancing both scientific understanding and practical applications of this multiferroic material.

\section{Materials and Methods:}
We used analytical-grade $Bi_2O_3$ and $Fe_2O_3$ powders obtained from Alfa Aesar as precursors, in stoichiometric ratios of 55$\colon$45 and 70$\colon$30 (mol \%), respectively. The respective regions in the phase diagram are indicated by the green arrows along the axis in Figure \ref{fig:phase_diagram}. The solid-state reaction method \cite{Valant2007Peculiarities,Orr2020sintering} was employed following established protocols. Initially, the precursor mixtures were manually ground for 30 minutes in an agate mortar using a pestle to ensure homogeneity. The crushed powders were then subjected to calcination in air at 400$^\circ$C for two hours. After calcination, approximately 600 mg of each powder sample was compacted under uniaxial pressure of 100 MPa to form pellets with an approximate diameter of 12.7 mm. Subsequently, these pellets were sintered in air at various temperatures ranging between 700$^\circ$C and 800$^\circ$C, each for a duration of two hours. The sintering time and maximum temperature were deliberately restricted to provide sufficient reaction time for $BiFeO_3$ formation, while minimizing the risk of decomposition into secondary phases or entry into the liquidus regime of the $Bi_2O_3$–$Fe_2O_3$ system, which destabilizes phase-pure $BiFeO_3$ in Bi-rich compositions. By maintaining our study within this thermal window, we ensured that the analysis focused on the crystal structure $R3c$ of $BiFeO_3$, with reduced risk of melt-mediated reactions that compromise phase purity.

Upon completion of the sintering process, all samples were quenched by placing them on an aluminium plate at room temperature, providing a rapid cooling path. X-ray diffraction (XRD) was conducted using a Rigaku SmartLab SE diffractometer with Cu-K$\alpha$ radiation ($\lambda$ = 1.5406 \AA) and an accuracy of $\Delta \theta = 0.02^\circ$. The estimation of the relative error in layer dimension came from the Bragg condition as $\frac{\Delta d}{d} = \cot \theta \cdot \Delta \theta$. Phase identification (crystal structure and corresponding lattice parameters) was carried out using peak fitting \cite{orr2021crystalline} and Rietveld refinement routines using the Profex software package \cite{Doebelin2015Profex}, with diffraction peaks compared with reference data from the Crystallography Open Database (COD) \cite{Grazulis2009Crystallography}. The phases identified included pure $BiFeO_3$ (BFO), sillenite-like structures, and mullite-like structures, with phase percentages calculated for each sintering temperature based on the fitted XRD peaks. The microstructural analysis of the samples was performed using a TESCAN Maia3 field-emission scanning electron microscope (FESEM). Raman data were acquired using an Oxford Instruments WITec alpha300 R Raman microscope system with an excitation wavelength of 532 nm.

\section{Results and Discussions}
\begin{figure}[htbp]
    \centering
    \includegraphics[width=0.95\textwidth]{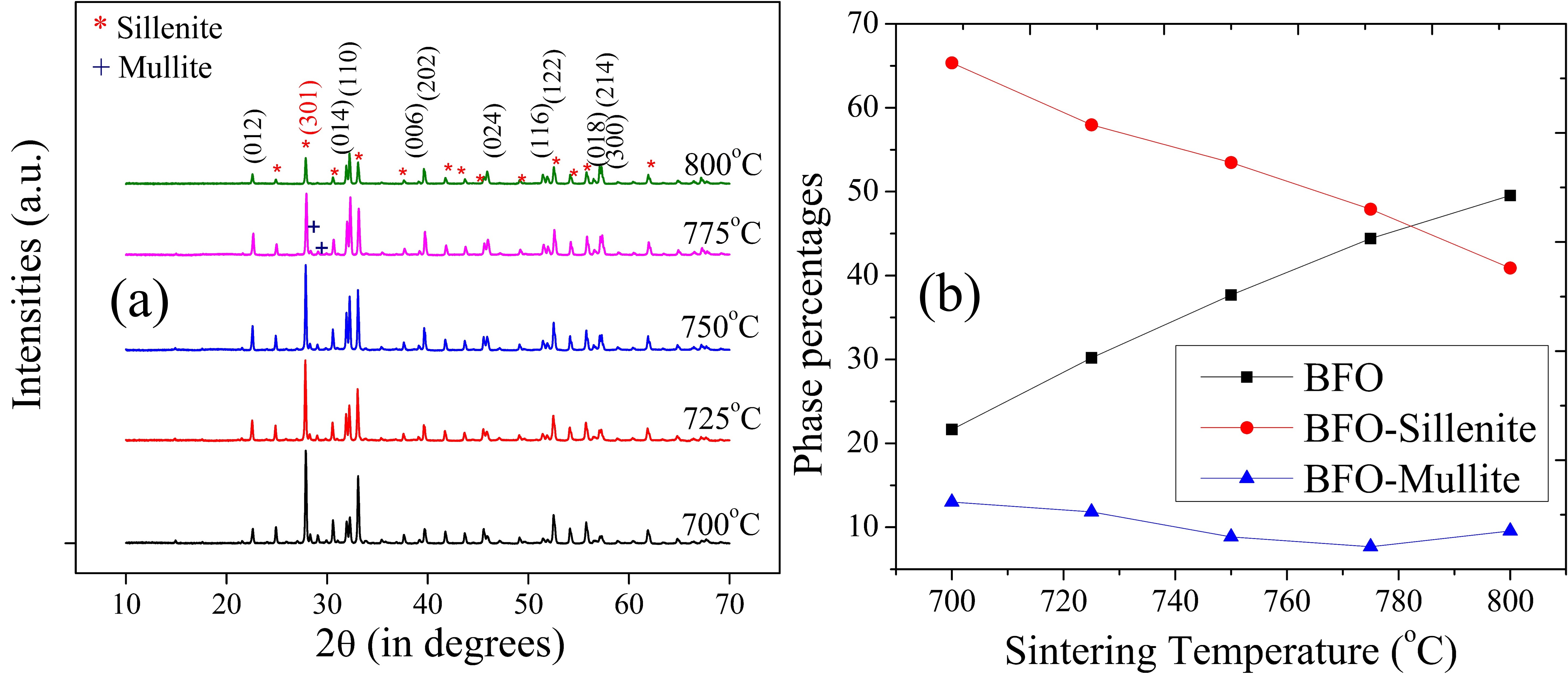}
    \caption{(a) XRD patterns from 55$\colon$45 (mol$\%$) samples sintered at different temperatures, with indexed phase-pure BFO (in black) and one prominent sillenite peak (in red) along with determined (b) corresponding phase percentages.}
    \label{fig:XRD_55_45}
\end{figure}

\begin{figure}[htbp]
    \centering
    \includegraphics[width=0.95\textwidth]{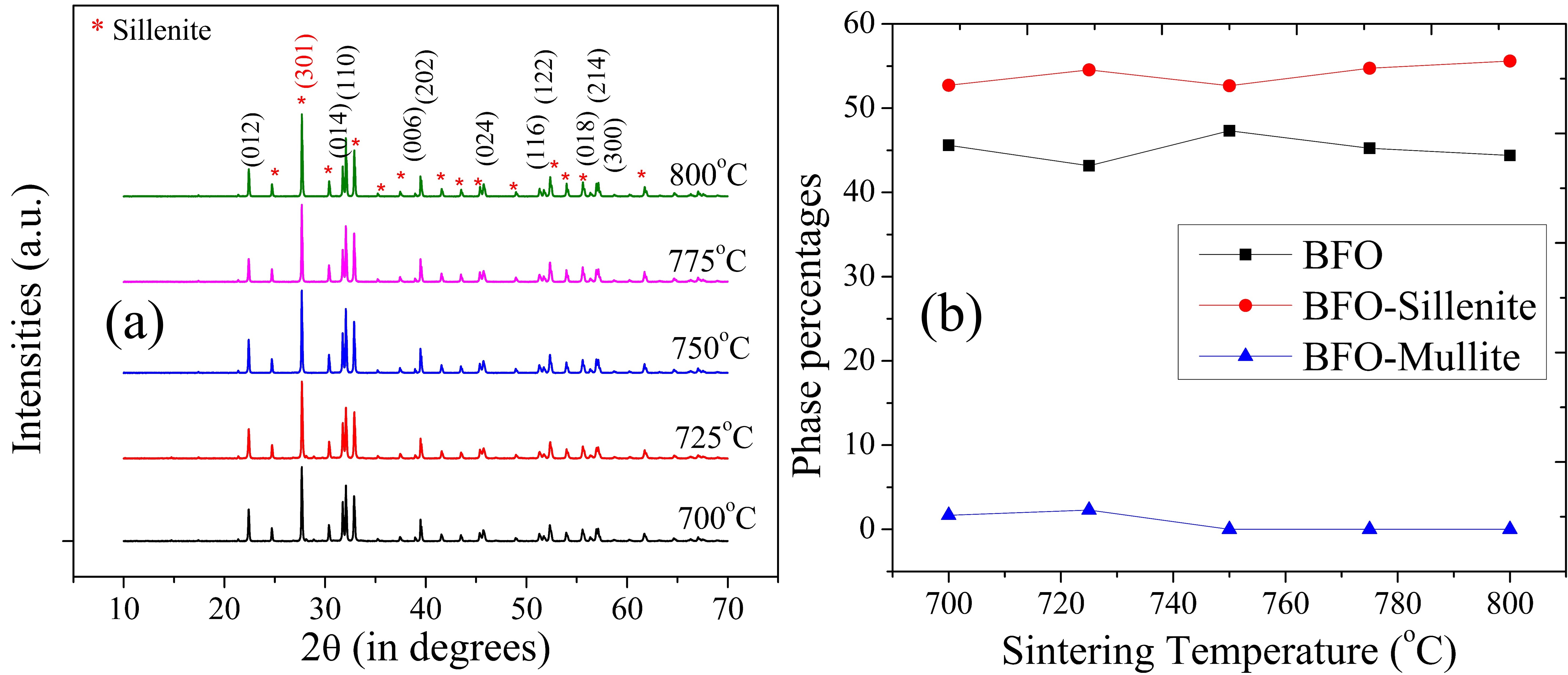}
    \caption{(a) XRD patterns from 70$\colon$30 (mol$\%$) samples sintered at different temperatures, with indexed phase-pure BFO (in black) and one prominent sillenite peak (in red) along with determined (b) corresponding phase percentages.}
    \label{fig:XRD_70_30}
\end{figure}

Figures \ref{fig:XRD_55_45}(a), \ref{fig:XRD_70_30}(a) provide a comprehensive overview of the X-ray diffraction (XRD) results, displaying the formation of both the primary phase $BiFeO_3$ and a range of secondary phases under all temperature conditions explored. The XRD patterns of the various samples reveal consistent peak positions, indicating similarities in their crystalline structures. However, in addition to the $BiFeO_3$ phase, secondary phases such as $Bi_{25}FeO_{39}$ (Sillenite-like structure) and $Bi_2Fe_4O_9$ (Mullite-like structure) are clearly present. These secondary phases vary in their molar or weight percentages depending on the specific temperature conditions, as detailed in Figure \ref{fig:XRD_55_45}(b), \ref{fig:XRD_70_30}(b). The results highlight how temperature influences the relative proportions of these phases, suggesting a complex interaction between the thermal conditions and the phase composition of the materials.

An important feature of this choice of temperatures is our observation that, in the 70$\colon$30 (mol\%) $Bi_2O_3-Fe_2O_3$ precursor mixture, a liquid phase emerges at $835^{\circ}C$. This behavior is documented in the phase diagram that we constructed in Figure \ref{fig:phase_diagram}, which integrates and compares several previous works \cite{Speranskaya1965Phase,Haumont2008Phase,Lu2011Phase, Palai2008Phase,Morozov2003Specific, Maitre2004Experimental, Achenbach1967Preparation}, further refined by our own experimental observation marked (g) in red. This melting is different from the well-known peritectic decomposition of $BiFeO_3$ at $~920^{\circ}C$ ($\beta\rightarrow\gamma$ transition) observed closer to the stoichiometric composition of 1$\colon$1. Identification of this onset of the liquid phase provides a direct rationale for the deliberate restriction of our sintering temperature to $800^{\circ}C$ in Section 2, ensuring that our study remained focused on the solid-state formation of $BiFeO_3$ while minimizing melt-assisted secondary phase formation.

The phase-pure $BiFeO_3$ exhibits a rhombohedrally distorted perovskite structure \cite{Orr2022Complex} under ambient conditions, which is categorized within the $R3c$ space group ($\alpha$-$BiFeO_3$ phase). In rhombohedral coordinates, its unit cell parameters are defined by $a_{rh}$ = 3.965 $\AA$ and $\alpha_{rh}$ = 89.3$^o$-89.4$^o$. Alternatively, the $BiFeO_3$ unit cell can also be described in terms of hexagonal symmetry, with lattice constants $a_{hex}$ = $b_{hex}$ = 5.58$\AA$ and $c_{hex}$ = 13.90$\AA$ \cite{Catalan2009Physics}.

This dual representation is evident from the appearance of two prominent characteristic peaks in X-ray diffraction (XRD) scans, corresponding to the crystal planes (104) and (110) of the R3c symmetry around 2$\theta$ = 32$^o$.

\begin{figure}[htbp]
    \centering
    \includegraphics[width=1.0\textwidth]{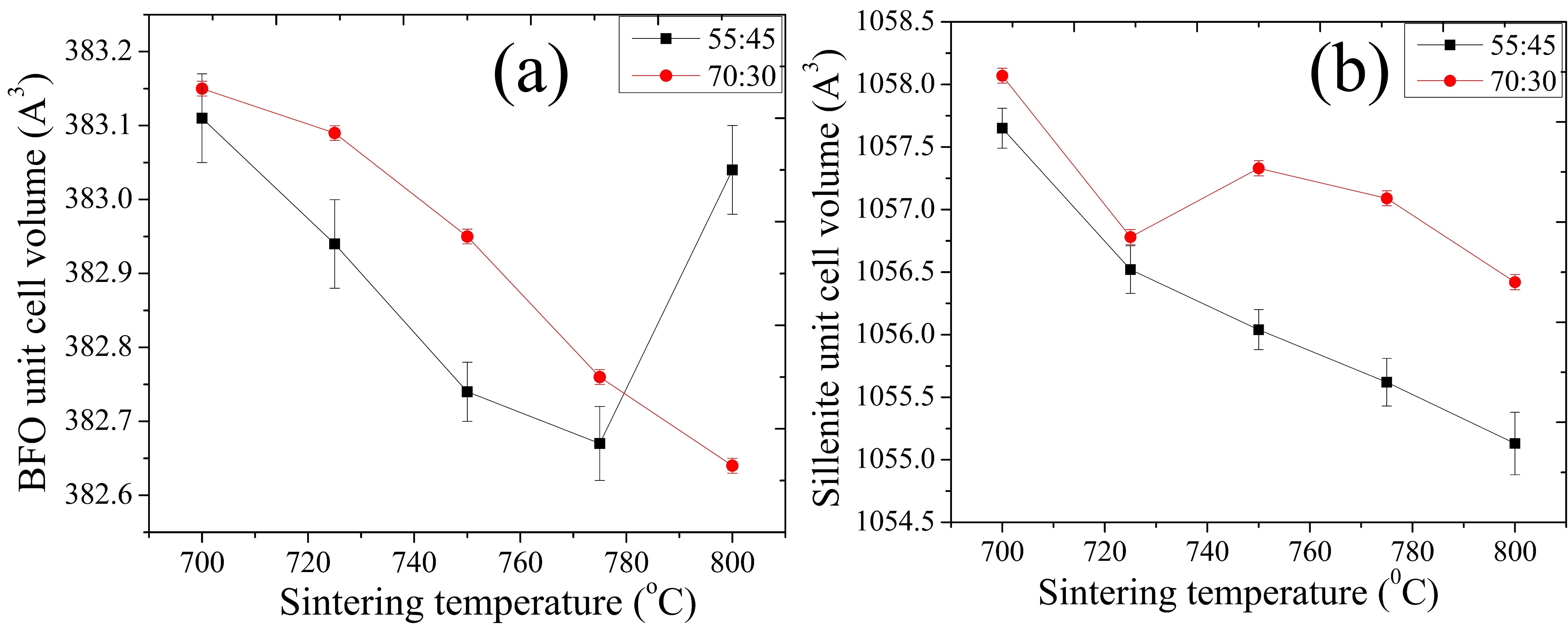}
    \caption{Calculated Unit cell volume of (a) pure phase Bismuth Ferrite (R3c) and (b) sillenite (I23) phase.}
    \label{fig:unit_cell_volume}
\end{figure}

In contrast, the secondary phases, Sillenite-like ($Bi_{25}FeO_{39}$) and Mullite-like \linebreak ($Bi_2Fe_4O_9$), adopt cubic (I23) and orthorhombic (Pbam) crystal structures, respectively. The identification of these phases was verified by comparing with data from the Crystallography Open Database (COD) \cite{Grazulis2009Crystallography}. The lattice parameters for these phases are obtained from the refined \cite{Doebelin2015Profex} peak positions corresponding to the crystal planes involved ((hkl) values) and the corresponding calculated unit cell volumes are illustrated in Figure \ref{fig:unit_cell_volume}. Since the presence of the Mullite-like phase is significantly lower in the samples compared to the other two phases (as shown in Figure \ref{fig:XRD_55_45}(b), \ref{fig:XRD_70_30}(b)), the focus of our discussion is limited to the phase-pure phase $BiFeO_3$ and the Sillenite-like phase.

Although a general trend of decreasing unit cell volumes with an increase in sintering temperature is observed for both precursor combinations, unit cell volumes are comparatively larger for samples with a higher concentration of bismuth in the precursor. This suggests that variations in the bismuth content significantly influence the structural properties of the resulting phases. The chemical reactions expected to occur during the synthesis of $BiFeO_3$ are as follows:

\begin{equation}
    Bi_2O_3 + Fe_2O_3 \longrightarrow 2BiFeO_3
\end{equation}

This reaction indicates that a 50$\colon$50 (mol$\%$) ratio of the precursors $Bi_2O_3$ and $Fe_2O_3$ is an ideal stoichiometric composition for the formation of phase-pure $BiFeO_3$. However, deviations from this, i.e., for the case of the 55$\colon$45 (mol$\%$) ratio, lead to an imbalance, where at this specific temperature stage the chemical reactivity of $Bi_2O_3$ increases \cite{Wesley2024, Piekiel2014}. This enhanced activity facilitates the formation reaction, described by the following equation:

\begin{equation}
    25Bi_2O_3 + Fe_2O_3 \longrightarrow 2Bi_{25}FeO_{39}
\end{equation}

Since more and more $Bi_2O_3$ are involved in the reaction due to their reactivity; the excess left-out $Fe_2O_3$, when it becomes more abundant, disrupts the formation of pure $BiFeO_3$ and leads to the iron-rich phase $Bi_2Fe_4O_9$ as an impurity. This reaction is illustrated as follows:

\begin{equation}
    Bi_2O_3 + 2Fe_2O_3 \longrightarrow Bi_2Fe_4O_9
\end{equation}

As the sintering temperature continues to increase, the bismuth-rich impurity phase $Bi_{25}FeO_{39}$ interacts with $Fe_2O_3$, leading to an enhancement in the proportion of the pure BFO phase. The resulting reaction can be expressed as follows \cite{Cheng2015Effect}:

\begin{equation}
    2Bi_{25}FeO_{39} + 24Fe_2O_3 \longrightarrow 50BiFeO_3
\end{equation}

Hence, there is a consistent increase in the proportion of phase-pure BFO with increasing sintering temperatures, as illustrated in Figure \ref{fig:XRD_55_45}(b). For samples with a 70$\colon$30 (mol$\%$) stoichiometric ratio of the precursor, as shown in Figure \ref{fig:XRD_70_30}(b), a similar reaction occurs. In this case a significant excess of Bismuth prevents the formation of the iron-rich impurity phase $Bi_2Fe_4O_9$, but promotes the bismuth-rich phases and the bismuth ferrite itself.

\begin{figure}[htbp]
    \centering
    \includegraphics[width=1.0\textwidth]{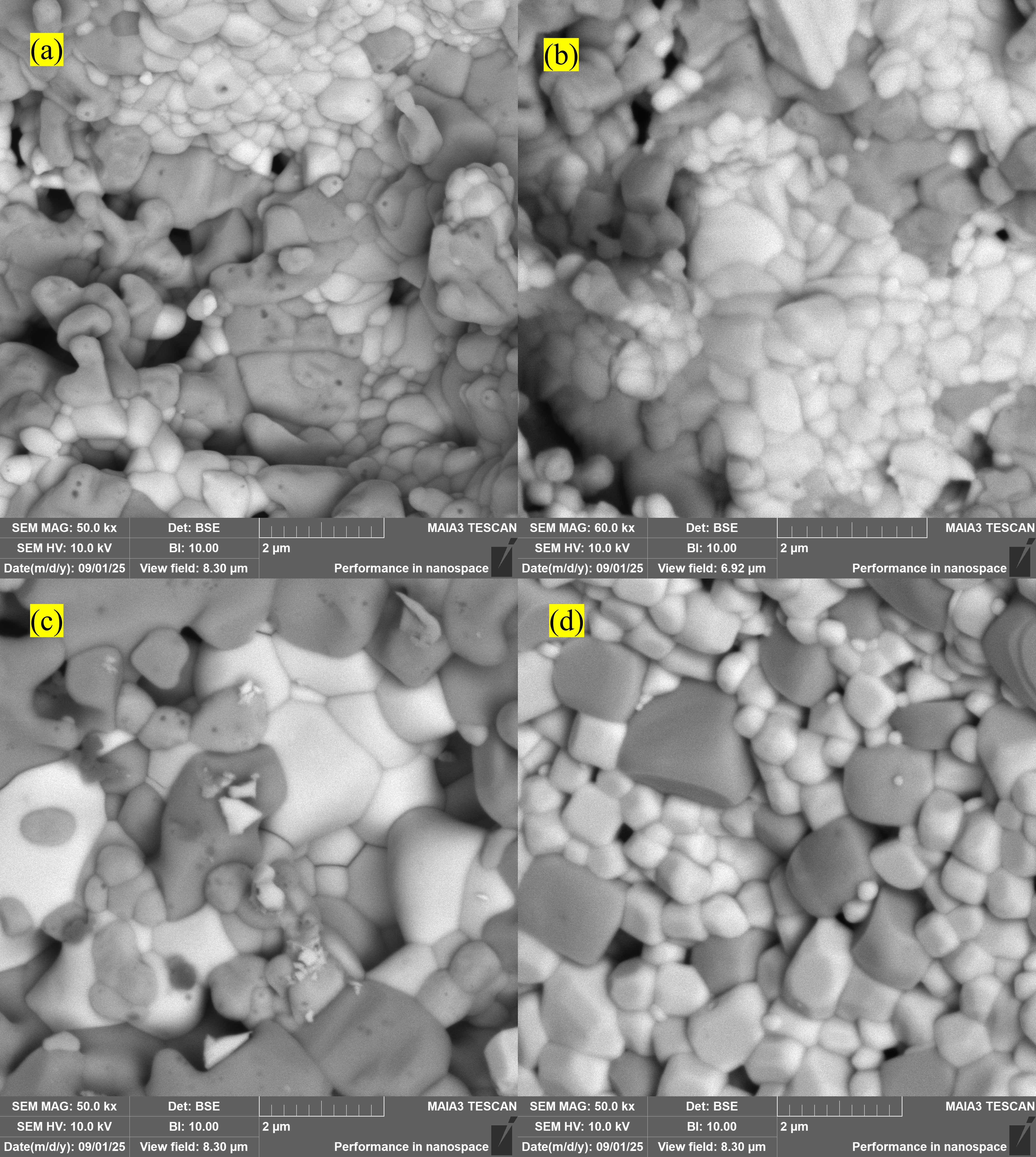}
    \caption{SEM micrographs (captured using backscattered electron) of (a) 55$\colon$45 (mol\%), (b) 70$\colon$30 (mol\%) samples sintered at 725$^\circ$C and (c) 55$\colon$45 (mol\%), (d) 70$\colon$30 (mol\%) samples sintered at 775$^\circ$C, respectively.}
    \label{fig:SEM_micrographs}
\end{figure}

SEM micrographs, captured using backscattered electrons presented in Figure \ref{fig:SEM_micrographs} clearly illustrate the presence of multiple phases as well as the presence of porous structures within the material. The pure bismuth ferrite phases are characterized by their distinctive flake-like morphology, featuring smooth, well-defined boundaries and appearing dark in the pictures. Sometimes these flakes also appear with slightly elongated grains. In contrast, the cubic-shaped structures, appearing whitish, observed in the micrographs are identified as Sillenite-like phases. Phase identification is also confirmed from the EDS data, added in the supplementary material.

The Crystallite sizes and microstrain for both phases, calculated using the Scherrer equation, are detailed in the Supplementary Material (Section 4.2). The analysis revealed that a sintering temperature of 775$^o$C predominantly promotes the growth of crystallite sizes, while minimizing micro-strain, which is typically introduced by grain boundaries in each phase. Additionally, it is evident from the results that the presence of moderate amounts of excess Bismuth contributes to the formation of larger crystallites, further enhancing the quality and purity of the BFO phase.

\begin{figure}[htbp]
    \centering
    \includegraphics[width=0.9\textwidth]{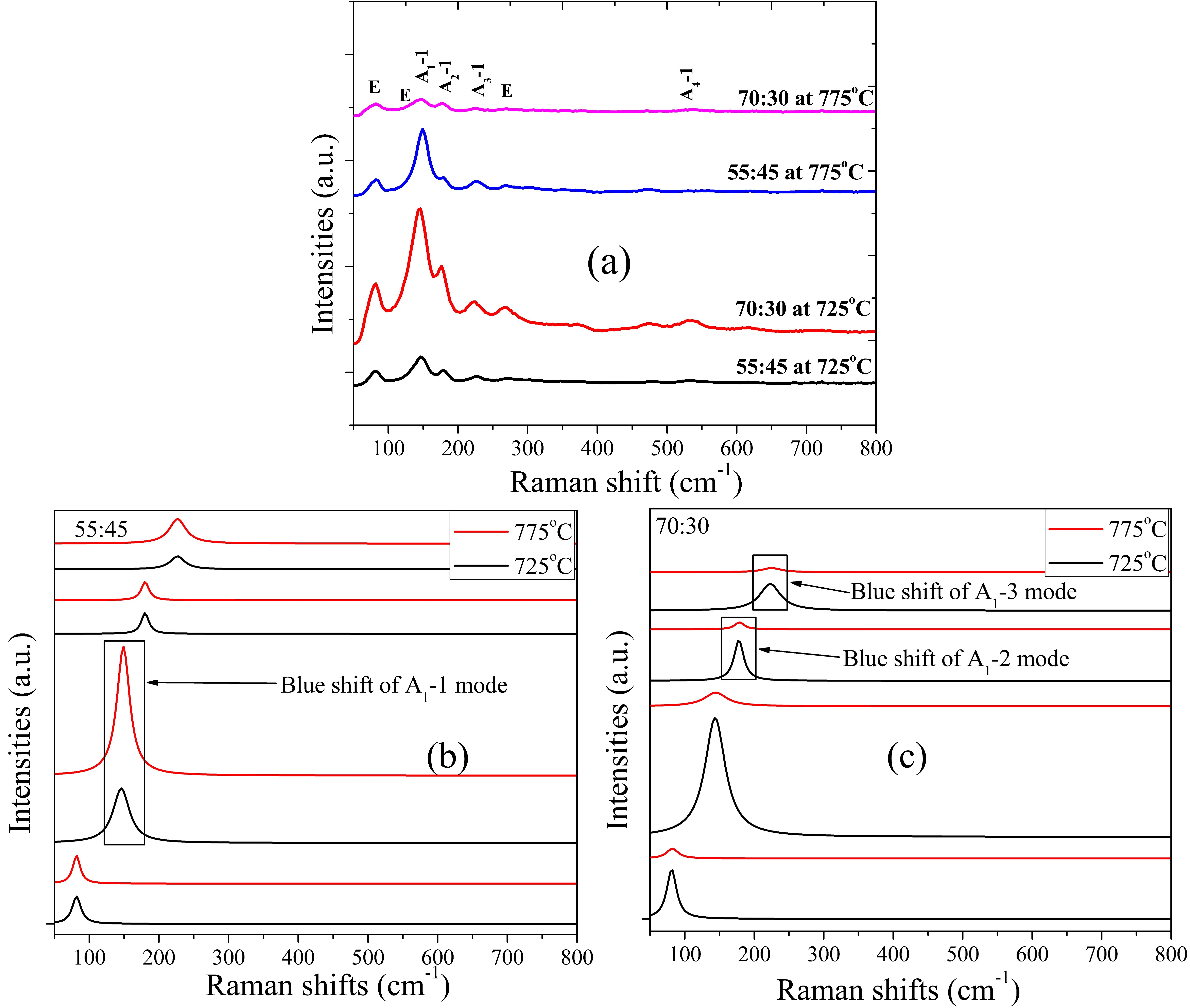}
    \caption{Micro-Raman analysis of $BiFeO_3$ phase: (a) spectra for all sample sets; deconvoluted spectra for (b) 55$\colon$45 samples and (c) 70$\colon$30 samples sintered at 725$^\circ$C and 775$^\circ$C respectively. Raman mode assignments are based on \cite{Jeon2011Enhanced}.}
    \label{fig:Raman_spectroscopy}
\end{figure}

Furthermore, Micro-Raman spectroscopy has elucidated distinct trends in the structural and vibrational properties of $BiFeO_3$ with varying bismuth stoichiometry (Figure \ref{fig:Raman_spectroscopy}). In samples with a 10$\%$ bismuth excess (55$\colon$45), the XRD results indicate unit cell contraction, while Raman spectroscopy reveals a blue shift and sharpening of the $A_1-1$ mode (Figure \ref{fig:Raman_spectroscopy} (c)), which is attributed to vibrations involving $Bi–O$ bonds \cite{Cazayous2007Electric}. These observations suggest improved crystallinity and stronger bonding in the lattice. Such enhancements are likely due to optimized stoichiometry, which leads to a more cohesive crystal structure.

In samples prepared with 40$\%$ excess Bismuth (70$\colon$30), distinct behavioural differences emerge compared to compositions with a lower bismuth content. Raman spectroscopy reveals significant alterations in the vibrational properties of these materials. Specifically, the $A_1-2$ and $A_1-3$ modes exhibit noticeable blue shifts, indicative of increased vibrational frequencies that may arise from strengthened lattice bonds or local compression\cite{Ouyang2008}. Concurrently, these peaks show pronounced broadening and marked reduction in intensity (Figure \ref{fig:Raman_spectroscopy}(d)). Such spectral changes suggest a defected structure with substantial strain and disorder in the polycrystalline structure. One possible source of such a disorder could be the presence of interstitial bismuth ions. However, X-ray diffraction (XRD) reveals that while unit cell dimensions shrink with higher sintering temperatures, bismuth-rich samples still exhibit larger cell volumes than those with lower bismuth content (Figure \ref{fig:unit_cell_volume}). Such spectral changes could suggest that the incorporation of surplus bismuth enhances certain bond strengths while simultaneously introducing substantial strain and disorder into the crystal structure, disrupting the uniformity of vibrational modes. One possible source for this anomaly could be the incorporation of $Bi^{3+}$ ions as defects, whose larger ionic radius (117 pm) compared to $Fe^{3+}$ (69 pm) \cite{ShannonRadii} induces significant lattice distortions. The dominant influence of $Bi^{3+}$—much like the "elephant in the room"— would overshadow the expected contraction trends. However, confirmation would require advanced tools like TEM or computational modelling.

These structural disruptions may require compensatory mechanisms within the lattice to mitigate the induced strain. One potential mechanism involves the partial reduction of $Fe^{3+}$ (69 pm) to $Fe^{2+}$ (92 pm) \cite{ShannonRadii}, as the larger ionic radius of $Fe^{2+}$ could alleviate lattice stress by better accommodating the expanded structure influenced by $Bi^{3+}$. This hypothesis is chemically feasible, particularly in bismuth-rich environments where local charge balance or oxygen stoichiometry might favour such a valence shift. However, this remains a speculative proposition at present, lacking direct experimental evidence such as X-ray photoelectron spectroscopy (XPS) or M\"{o}ssbauer spectroscopy to confirm the presence of $Fe^{2+}$. Further investigation is warranted to validate this mechanism and explore alternative compensatory processes, such as the formation of oxygen vacancies or phase segregation, which could similarly address the observed lattice strain.

\begin{figure}[htbp]
\centering
\includegraphics[width=\textwidth]{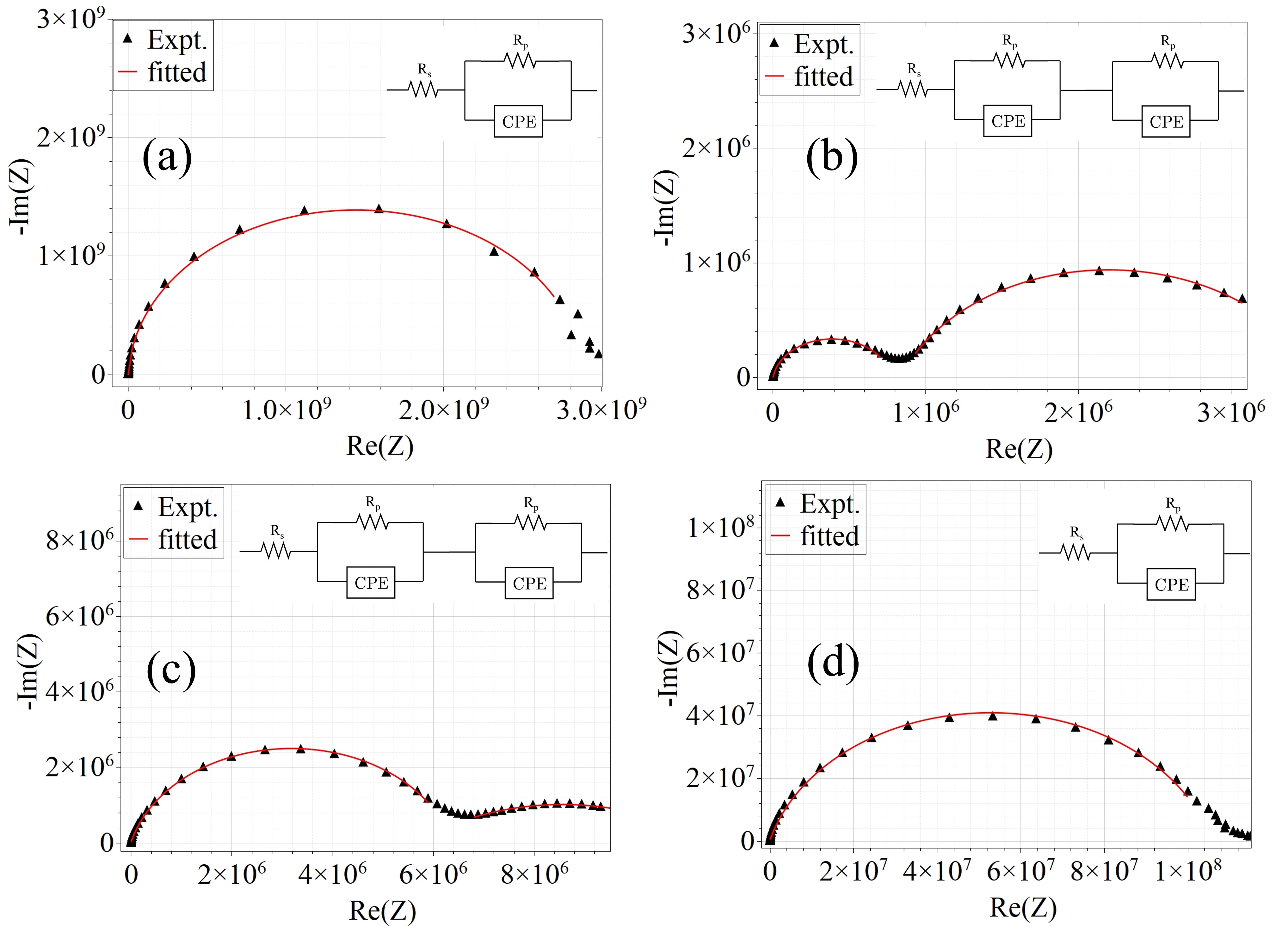}
\caption{Combined Nyquist plots of BiFeO$_3$ ceramics: (a) 55$\colon$45 at 725$^o$C, (b) 55$\colon$45 at 775$^o$C, (c) 70$\colon$30 at 725$^o$C, and (d) 70$\colon$30 at 775$^o$C. Symbols represent experimental data; solid lines are Levenberg–Marquardt fits for semicircles.}
\label{fig:Nyquist_All}
\end{figure}

In addition to structural, morphological and vibrational characterizations, Nyquist plot analysis was performed by impedance spectroscopy on BiFeO$_3$ samples with precursor ratios of 55$\colon$45 and 70$\colon$30 mol\% Bi$_2$O$_3$ : Fe$_2$O$_3$, sintered at 725$^o$C and 775$^o$C. This technique involves plotting the negative imaginary impedance (–Im(Z)) against the real impedance (Re(Z)) over a broad frequency range to resolve electrical relaxation processes (Figure~\ref{fig:Nyquist_All}). Each plot shows one or two depressed semicircular arcs, attributed to grain (bulk) and grain boundary contributions. The experimental data were fitted using the Levenberg–Marquardt algorithm in QtiPlot  (convergence tolerance: $10^{-9}$) to the semicircle equation: $(Z' - h)^2 + (Z'' - k)^2 = r^2$, where $(h, k)$ is the center, $r$ is the radius and $f_{\text{max}}$ is the frequency at the maximum of the arc. These fitted parameters are reported in Table~\ref{Tab:CircleParams}.

\begin{table}[htbp]
\centering
\caption{Fitted circle parameters from Nyquist analysis.}
\label{Tab:CircleParams}
\begin{tabular}{|c|c|c|c|c|c|}
\hline
\textbf{Sample} & \textbf{Arc} & \textbf{$h$ ($\Omega$)} & \textbf{$k$ ($\Omega$)} & \textbf{$r$ ($\Omega$)} & \textbf{$f_{\max}$ (Hz)} \\
\hline
\hline
55$\colon$45 at 725$^o$C & 1 & 1.44254 $\times 10^{9}$ & $-$5.46 $\times 10^{7}$ & 1.44335 $\times 10^{9}$ & 2.6561 \\
\hline
55$\colon$45 at 775$^o$C & 1 & 3.88 $\times 10^{5}$ & $-$5.58 $\times 10^{4}$ & 3.91 $\times 10^{5}$ & 1587.3 \\
\hline
55$\colon$45 at 775$^o$C & 2 & 2.20 $\times 10^{6}$ & $-$5.23 $\times 10^{5}$ & 1.46 $\times 10^{6}$ & 0.49385 \\
\hline
70$\colon$30 at 725$^o$C & 1 & 3.18 $\times 10^{6}$ & $-$7.76 $\times 10^{5}$ & 3.28 $\times 10^{6}$ & 413.2 \\
\hline
70$\colon$30 at 725$^o$C & 2 & 8.56 $\times 10^{6}$ & $-$3.69 $\times 10^{6}$ & 4.71 $\times 10^{6}$ & 0.25197 \\
\hline
70$\colon$30 at 775$^o$C & 1 & 5.28 $\times 10^{7}$ & $-$1.37 $\times 10^{7}$ & 5.47 $\times 10^{7}$ & 76.828 \\
\hline
\end{tabular}
\end{table}

Key RC parameters in the equivalent circuit model include the parallel resistance of each arc 
($R_p = 2r$), the relaxation time ($\tau = 1/(2\pi f_{\max})$), and the effective capacitance 
($C_{\text{eff}} = \tau / R_p = 1/(2\pi f_{\max} R_p)$), while the semicircle centres 
$(h,k)$ yield the series resistance ($R_s = h - r$). Since all semicircles are depressed 
($k < 0$, centre below the real axis), the degree of depression was quantified using the 
geometric angle
$\phi = \arcsin\left(\frac{|k|}{r}\right)$,
which directly relates to the Cole-Cole stretch parameter. The corresponding constant-phase-element (CPE) exponent was then calculated as $n = 1 - \frac{2\phi}{\pi}$ (or equivalently $n = 1 - 2\phi/180^\circ$ when $\phi$ is in degrees), ensuring full consistency with a non-ideal representation $(R_p \parallel \text{CPE})$ \cite{BarsoukovMacdonald2018}. For completeness, the CPE prefactor $Q$ was determined by $Q = \frac{1}{R_p \, (2\pi f_{\max})^n}$. From the high-frequency arc, we extracted the resistance representing the intrinsic bulk response of the material (minimal transient polarization contributions). We therefore report an effective bulk AC conductivity $\sigma_{\text{ac,bulk}} = \frac{t}{A \, R_{\text{bulk}}}$, where $R_{\text{bulk}} = R_p$ of the first (high-frequency) arc, $t$ is the pellet thickness, and $A$ is the electrode area. All derived quantities are summarized in Table~\ref{tab:electrical_params}.

\begin{table}[htbp]
\centering
\caption{Complete electrical parameters extracted from Nyquist analysis using parallel (R$\parallel$ CPE) equivalent circuits.}
\label{tab:electrical_params}
\begin{tabular}{|c|c|c|c|c|c|c|c|}
\hline
\textbf{Sample}            & \textbf{Arc} & \textbf{$Q$ (S$\cdot$s$^n$)} & \textbf{$C_{\text{eff}}$ (F)} & \textbf{$\phi$ ($^\circ$)} & \textbf{$n$} & \textbf{$\tau$ (s)} & \textbf{$\sigma_{\text{ac,bulk}}$ (S/m)} \\
\hline
\hline
55$\colon$45 at 725$^\circ$C     & 1 & 2.22$\times$10$^{-11}$ & 2.08$\times$10$^{-11}$ & 2.17  & 0.976 & 0.059921 & 3.12$\times$10$^{-8}$ \\
\hline
55$\colon$45 at 775$^\circ$C     & 1 & 2.96$\times$10$^{-10}$ & 1.28$\times$10$^{-10}$ & 8.20  & 0.909 & 0.000100 & 7.81$\times$10$^{-5}$ \\
\hline
55$\colon$45 at 775$^\circ$C     & 2 & 1.44$\times$10$^{-7}$  & 1.10$\times$10$^{-7}$  & 20.99 & 0.767 & 0.322274 & — \\
\hline
70$\colon$30 at 725$^\circ$C     & 1 & 1.94$\times$10$^{-10}$ & 5.87$\times$10$^{-11}$ & 13.69 & 0.848 & 0.000385 & 5.24$\times$10$^{-6}$ \\
\hline
70$\colon$30 at 725$^\circ$C     & 2 & 8.72$\times$10$^{-8}$  & 6.71$\times$10$^{-8}$  & 51.58 & 0.427 & 0.631642 & — \\
\hline
70$\colon$30 at 775$^\circ$C     & 1 & 5.12$\times$10$^{-11}$ & 1.89$\times$10$^{-11}$ & 14.50 & 0.839 & 0.002072 & 5.58$\times$10$^{-7}$ \\
\hline
\end{tabular}
\end{table}

The electrical microstructure resolved through Nyquist analysis closely reflects the structural refinement identified from XRD, SEM and Raman spectroscopy, with clear dependencies on both precursor stoichiometry and sintering temperature. For the 55$\colon$45 composition, increasing the sintering temperature from 725\,$^\circ$C to 775\,$^\circ$C results in a dramatic collapse of the bulk resistance, from $R_p = 2.89 \times 10^{9}$\,$\Omega$ to $7.82 \times 10^{5}$\,$\Omega$, accompanied by a similarly reduced series resistance. This improvement indicates the removal of blocking Bi$_{25}$FeO$_{39}$ domains and the formation of well-connected, phase-pure BiFeO$_3$ grains. The corresponding dielectric response exhibits a near-Debye CPE exponent ($n = 0.909$), a small prefactor ($Q \approx 3 \times 10^{-10}$\,S$\cdot$s$^n$), and a reduced effective capacitance ($C_{\text{eff}} \approx 10^{-10}$\,F), all of which are indication of homogeneous, low-defect grain interiors. The resulting enhancement in bulk AC conductivity—from $3.12 \times 10^{-8}$ to $7.81 \times 10^{-5}$\,S\,m$^{-1}$—correlates strongly with the lattice contraction and Raman $A_1$-mode sharpening that signify reduced strain and strengthened Bi--O bonding.

At 725\,$^\circ$C, the same composition exhibits a massive semicircle (radius $\sim 1.44 \times 10^{9}$\,$\Omega$), reflecting the persistence of insulating impurity phases that limit charge percolation, even though the CPE exponent remains high ($n = 0.976$). Upon sintering at 775\,$^\circ$C, a second, low-frequency arc appears with $n = 0.767$ and $C_{\text{eff}} \sim 10^{-7}$\,F, probably because of a thin Bi$_2$O$_3$-rich intergranular film formed during liquid-phase-assisted densification; however, this contribution is minor relative to the dominant bulk relaxation.

In contrast, the Bi-rich 70$\colon$30 ceramics exhibit persistent structural disorder across both sintering temperatures. At 725\,$^\circ$C, the low-frequency arc is highly depressed ($n = 0.427$, $\phi > 50^\circ$) and accompanied by a large CPE prefactor ($Q \approx 8.7 \times 10^{-8}$\,S$\cdot$s$^n$), indicating a broad distribution of relaxation times associated with defective, compositionally inhomogeneous grain boundaries enriched in Bi$^{3+}$ antisites and Fe-oxidised species—consistent with their expanded lattice parameters and broadened Raman modes. Even after sintering at 775\,$^\circ$C, the impedance response remains dominated by a merged, resistive arc with low conductivity ($5.58 \times 10^{-7}$\,S\,m$^{-1}$), confirming that excessive Bi$_2$O$_3$ stabilises insulating secondary phases and inhibits microstructural homogenisation.

Overall, the evolution from highly depressed, multi-arc Nyquist profiles with low CPE exponents to single, near-Debye semicircles with $n > 0.9$ provides a quantitative electrical signature of increasing structural order. Only the slightly Bi-rich 55$\colon$45 precursor composition sintered at 775\,$^\circ$C yields phase-pure, well-crystallised BiFeO$_3$ with minimal grain-boundary barriers and enhanced charge-carrier mobility, in complete agreement with the trends established by XRD lattice contraction and Raman mode sharpening.
The discussed dielectric measurements were conducted on quenched samples. Following this work we will conduct a study with the same compositions at high temperatures using a time-domain measurement system adapted for ceramics \cite{orr2018high}.

\section{Conclusion}
A phase-diagram-guided synthesis strategy was employed to tailor the structural and dielectric behaviour of BiFeO$_3$ ceramics within the Bi$_2$O$_3$–Fe$_2$O$_3$ system. Refinement of the Bi-rich region of the phase diagram identified the onset of liquid-phase formation near 835$^\circ$C. Solid-state synthesis using non-stoichiometric precursor ratios demonstrated that the 55$\colon$45 (Bi$_2$O$_3$ $\colon$ Fe$_2$O$_3$) composition sintered at 775$^\circ$C yields phase-pure rhombohedral $R3c$ BiFeO$_3$ with lattice contraction, suppressed sillenite- and mullite-type impurities, and homogeneous grain development. Impedance spectroscopy revealed a dominant bulk semicircle with minimal series resistance, low bulk resistance, and a near-Debye dielectric response (n $\approx$ 0.9), accompanied by a four-orders-of-magnitude increase in electrical conductivity relative to the low-temperature state. In contrast, Bi-rich (70$\colon$30) compositions exhibited persistent secondary phases, enlarged $R_p$ values, and strongly depressed relaxation behaviour, indicating defect-dominated charge transport and resistive grain boundaries. These results establish a direct processing–structure–property pathway demonstrating that precise stoichiometry and thermal control enable high-purity BiFeO$_3$ with stable dielectric performance. The optimal 55$\colon$45 precursor sintered at 775$^\circ$C provides a reproducible route for engineering low-loss, high-mobility electrical behaviour, supporting the development of BFO-based multiferroic and energy-conversion technologies.

	\bibliographystyle{unsrt} 
	\bibliography{references} 


\section*{Supplementary Material}

\subsection{Calculation of Unit Cell Volume and Uncertainty from Lattice Parameters}
\label{sec:S1}
The unit cell volume of \textbf{BiFeO$_3$} (R3c, hexagonal setting) is calculated as 
$V = \frac{\sqrt{3}}{2} a^2 c$ 
while for \textbf{Bi$_2$5FeO$_{39}$} (I23, cubic), 
$V = a^3$. 
The uncertainty  in volume propagates using standard error analysis: for R3c, the relative error is
$\frac{\Delta V}{V} = 2\frac{\Delta a}{a} + \frac{\Delta c}{c}$;  
for I23, 
$\frac{\Delta V}{V} = 3\frac{\Delta a}{a}$, 
with the absolute error 
$\Delta V = V \times \frac{\Delta V}{V}$. 
All calculations assume uncorrelated, normally distributed lattice parameter errors from Rietveld refinement.

\subsection{Crystallite Size and Microstrain Analysis}
\label{sec:S2}
To gain a deeper understanding of these results, we calculated the crystallite sizes (D) as shown in Table \ref{Tab:cryst_55_45}-\ref{Tab:cryst_70_30}, using the Scherrer equation \cite{Patterson1939Scherrer}:

\begin{equation}
    D= \frac{K \lambda}{\beta \sin \theta}
\end{equation}

Whereas, the microstrain ($\varepsilon$) corresponding to each prominent peak in XRD scans were estimated using the following formula \cite{Williamson1953XRay}:

\begin{equation}
    \varepsilon = \frac{\beta}{4 \tan \theta}
\end{equation}

Here, $\beta$ stands for the full width at half maximum (FWHM) of the X-ray diffraction (XRD) peak measured at the diffraction angle 2$\theta$, $\lambda$ is the wavelength of the X-ray, and K is the shape factor. Error in Crystallite Size ($\Delta$D) is calculated using the relation: $\Delta D = D \cdot \sqrt{(\frac{\Delta \beta}{\beta})^2 + (\tan \theta \cdot \Delta \theta)^2}$. This accounts for uncertainties in the peak broadening ($\beta$) and angular measurements ($\theta$), while the error in Microstrain ($\Delta \varepsilon$) is determined by: $\Delta \varepsilon = \varepsilon \cdot \sqrt{(\frac{\Delta \beta}{\beta})^2 + (\cot \theta \cdot \Delta \theta)^2}$. This captures the contributions from instrumental resolution and angular deviations, indicating error propagation in both the parameters.

\begin{table}[htbp]
    \centering
    \caption{\label{Tab:cryst_55_45}Calculated crystallite size and micro-strain of 55$\colon$45 (mol\%) samples using Scherrer equation from the most prominent peak in XRD data.}
    \resizebox{\textwidth}{!}{%
    \begin{tabular}{|c|c|c|c|c|c|c|}
        \hline
        \textbf{Sintering Temp. ($^\circ$C)} & \multicolumn{3}{c|}{\textbf{BiFeO$_3$}} & \multicolumn{3}{c|}{\textbf{Bi$_{25}$FeO$_{39}$}} \\
        \hline
        & 2$\theta$ (110) & Cryst. Size (nm) & Micro-Strain ($\varepsilon \times 10^{-4}$) & 2$\theta$ (301) & Cryst. Size (nm) & Micro-Strain ($\varepsilon \times 10^{-4}$) \\
        \hline
        700 & 32.26 & 100.70 $\pm$ 2.14 & 6.52 $\pm$ 1.38 & 27.89 & 114.47 $\pm$ 2.88 & 6.55 $\pm$ 1.65 \\
        725 & 32.21 & 88.77 $\pm$ 1.66  & 7.41 $\pm$ 1.39 & 27.83 & 88.69 $\pm$ 1.73  & 8.46 $\pm$ 1.65 \\
        750 & 32.24 & 95.60 $\pm$ 1.92  & 6.87 $\pm$ 1.38 & 27.87 & 91.85 $\pm$ 1.86  & 8.16 $\pm$ 1.65 \\
        775 & 32.34 & 201.71 $\pm$ 8.56 & 3.25 $\pm$ 1.38 & 27.94 & 99.97 $\pm$ 2.20  & 7.48 $\pm$ 1.65 \\
        800 & 32.22 & 185.13 $\pm$ 7.22 & 3.55 $\pm$ 1.38 & 27.88 & 231.84 $\pm$ 11.83 & 3.23 $\pm$ 1.65 \\
        \hline
    \end{tabular}%
    }
\end{table}

\begin{table}[htbp]
    \centering
    \caption{\label{Tab:cryst_70_30}Calculated crystallite size and micro-strain of 70$\colon$30 (mol\%) samples using Scherrer equation from the most prominent peak in XRD data.}
    \resizebox{\textwidth}{!}{%
    \begin{tabular}{|c|c|c|c|c|c|c|}
        \hline
        \textbf{Sintering Temp. ($^\circ$C)} & \multicolumn{3}{c|}{\textbf{BiFeO$_3$}} & \multicolumn{3}{c|}{\textbf{Bi$_{25}$FeO$_{39}$}} \\
        \hline
        & 2$\theta$ (110) & Cryst. Size (nm) & Micro-Strain ($\varepsilon \times 10^{-4}$) & 2$\theta$ (301) & Cryst. Size (nm) & Micro-Strain ($\varepsilon \times 10^{-4}$) \\
        \hline
        700 & 32.08 & 117.25 $\pm$ 2.90 & 5.63 $\pm$ 1.39 & 27.69 & 169.45 $\pm$ 6.33 & 4.45 $\pm$ 1.66 \\
        725 & 32.08 & 108.92 $\pm$ 2.50 & 6.06 $\pm$ 1.39 & 27.69 & 174.59 $\pm$ 6.72 & 4.32 $\pm$ 1.66 \\
        750 & 32.07 & 140.31 $\pm$ 4.15 & 4.70 $\pm$ 1.39 & 27.68 & 198.55 $\pm$ 8.69 & 3.80 $\pm$ 1.66 \\
        775 & 32.07 & 140.59 $\pm$ 4.17 & 4.70 $\pm$ 1.39 & 27.68 & 241.68 $\pm$ 12.88 & 3.12 $\pm$ 1.66 \\
        800 & 32.08 & 117.25 $\pm$ 2.90 & 5.63 $\pm$ 1.39 & 27.69 & 169.45 $\pm$ 6.33 & 4.45 $\pm$ 1.66 \\
        \hline
    \end{tabular}%
    }
\end{table}

\subsection{SEM-EDS Analysis of Phase Coexistence}
\label{sec:S3}
\begin{figure}[htbp]
    \centering
    \includegraphics[width=1.1\textwidth]{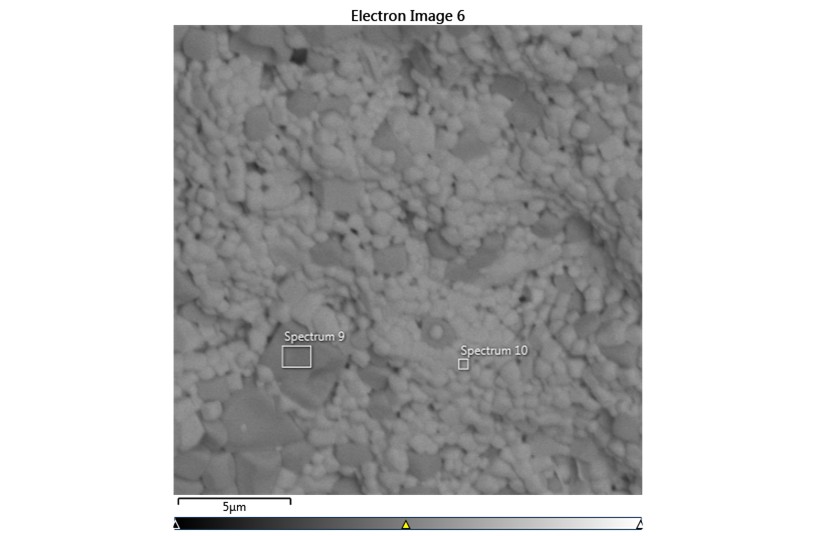}
    \caption{Backscattered electron (BSE) SEM micrograph showing BiFeO$_3$ (dark, Spectrum 9) and sillenite-type phase (bright, Spectrum 10).}
    \label{fig:sem_bse}
\end{figure}

To confirm phase coexistence in Bi-rich samples, a backscattered electron (BSE) SEM micrograph was obtained (Figure \ref{fig:sem_bse}), where contrast highlights different phases: the darker region corresponds to BiFeO$_3$ (marked as Spectrum 9) and the brighter region to the sillenite-type $Bi_{25}FeO_{39}$ phase (marked as Spectrum 10). 

\begin{figure}[htbp]
    \centering
    \includegraphics[width=0.9\textwidth]{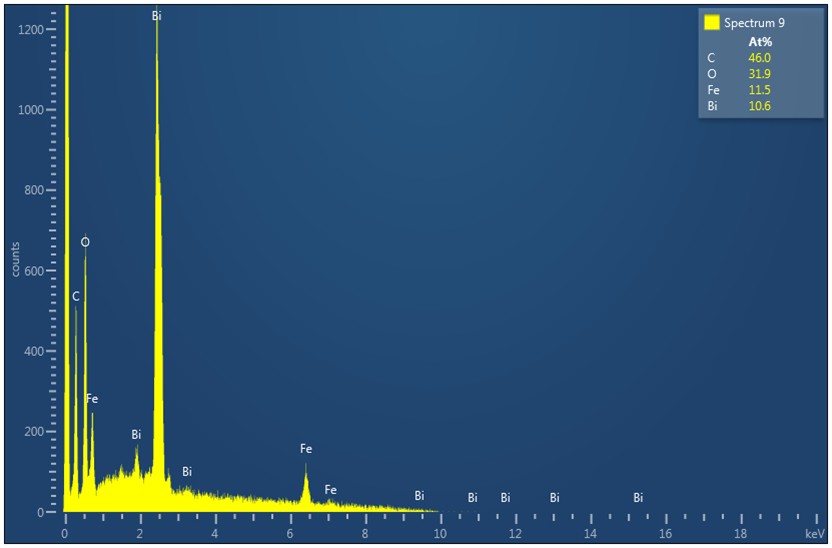}
    \caption{EDS spectrum corresponding to Spectrum 9 (dark region in Figure \ref{fig:sem_bse}), indicating composition close to BiFeO$_3$., i.e. comparable At\% of Bi, Fe}
    \label{fig:eds9}
\end{figure}

\begin{figure}[htbp]
    \centering
    \includegraphics[width=0.9\textwidth]{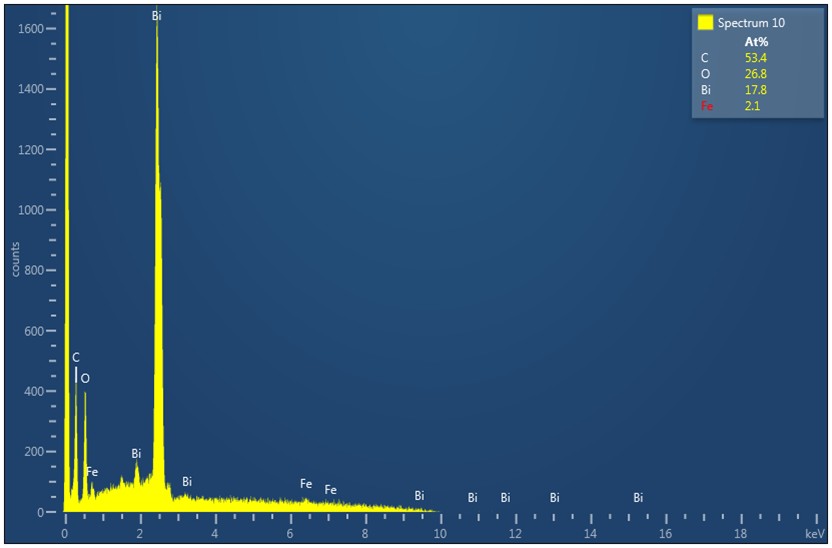}
    \caption{EDS spectrum corresponding to Spectrum 10 (bright region in Figure \ref{fig:sem_bse}), indicating Bi-rich composition similar sillenite-type $Bi_{25}FeO_{39}$.}
    \label{fig:eds10}
\end{figure}

Energy-dispersive spectroscopy (EDS) spectra for these regions are shown in Figure \ref{fig:eds9} and Figure \ref{fig:eds10}. The relative atomic percentages of Bi and Fe obtained from the EDS indices are consistent with the assignment of BiFeO$_3$ in Spectrum 9 and Bi-rich sillenite phase in Spectrum 10.

\end{document}